\documentclass[10pt,twocolumn,aps,prx,superscriptaddress,floatfix]{revtex4}

\usepackage{epsfig}
\usepackage{amsmath, amssymb, amsfonts, mathrsfs}
\usepackage{amsthm}
\usepackage{graphicx}
\usepackage{color}
\usepackage{xcolor}
\usepackage{hyperref}
\hypersetup{
    colorlinks,
    linkcolor={rgb:red,2; blue,1},
    citecolor={blue!60!black},
    urlcolor={blue!60!black}
}
\usepackage{bbm}
\usepackage{soul}

\newcommand{\ket}[1]{\vert{#1}\rangle} 
\newcommand{\bra}[1]{\langle{#1}\vert} 
 
\DeclareMathOperator{\Tr}{Tr}

\newcommand{\beq}{\begin{equation}}
\newcommand{\eeq}{\end{equation}}

\newcommand{\be}{\begin{equation}}
\newcommand{\ee}{\end{equation}}

\newcommand{\ben}{\begin{eqnarray}}
\newcommand{\een}{\end{eqnarray}}
\usepackage{amsmath}

\renewcommand{\dag}{\dagger}
\newcommand{\e}{\mathrm{e}}
\newcommand{\LL}{\mathcal{L}}
\newcommand{\DD}{\mathcal{D}}
\newcommand{\HH}{\mathcal{H}}
\newcommand{\PP}{\mathcal{P}}
\newcommand{\dt}[1]{\frac{\mathrm{d}#1}{\mathrm{d}t}}
\newcommand{\ii}{\mathrm{i}}
\newcommand{\dd}{\mathrm{d}}

\theoremstyle{definition}

\begin{document}

\title{Autonomous quantum clocks: does thermodynamics limit our ability to measure time?}

\author{Paul Erker$^*$}
\affiliation{Universitat Autonoma de Barcelona, 08193 Bellaterra, Barcelona, Spain}
\affiliation{Faculty of Informatics, Universit\`{a} della Svizzera italiana, Via G. Buffi 13, 6900 Lugano, Switzerland}
%\affiliation{Facolt\`{a} indipendente di Gandria, Lunga scala, 6978 Gandria, Switzerland}
\author{Mark T. Mitchison$^*$}
\affiliation{Quantum Optics and Laser Science Group, Blackett Laboratory, Imperial College London, London SW7 2BW, United Kingdom}
\affiliation{Institut f\"{u}r Theoretische Physik, Albert-Einstein Allee 11, Universit\"{a}t Ulm, 89069 Ulm, Germany}
\author{Ralph Silva$^*$}
\affiliation{Group of Applied Physics, University of Geneva, 1211 Geneva 4, Switzerland}
\author{Mischa P. Woods}
%\affiliation{Centre for Quantum Technologies, National University of Singapore, 3 Science Drive 2, Singapore 117543}
\affiliation{University College London, Department of Physics \& Astronomy, London WC1E 6BT, United Kingdom}
\affiliation{QuTech, Delft University of Technology, Lorentzweg 1, 2611 CJ Delft, Netherlands}
\author{Nicolas Brunner}
\affiliation{Group of Applied Physics, University of Geneva, 1211 Geneva 4, Switzerland}
\author{Marcus Huber}
\affiliation{Institute for Quantum Optics and Quantum Information (IQOQI), Austrian Academy of Sciences, %Boltzmanngasse 3, 
A-1090 Vienna, Austria\\$^*$\text{these authors contributed equally to this paper}}

\date{\today}

\begin{abstract}
Time remains one of the least well understood concepts in physics, most notably in quantum mechanics. A central goal is to find the fundamental limits of measuring time. One of the main obstacles is the fact that time is not an observable and thus has to be measured indirectly.  Here we explore these questions by introducing a model of time measurements that is complete and autonomous. Specifically, our autonomous quantum clock consists of a system out of thermal equilibrium --- a prerequisite for any system to function as a clock --- powered by minimal resources, namely two thermal baths at different temperatures. Through a detailed analysis of this specific clock model, we find that the laws of thermodynamics dictate a trade-off between the amount of dissipated heat and the clock's performance in terms of its accuracy and resolution. Our results furthermore imply that a fundamental entropy production is associated with the operation of \textit{any} autonomous quantum clock, assuming that quantum machines cannot achieve perfect efficiency at finite power. More generally, autonomous clocks provide a natural framework for the exploration of fundamental questions about time in quantum theory and beyond.
\end{abstract}

\maketitle

\section{Introduction}
Although quantum systems provide the most accurate measurements of time \cite{clock1,clock2,clock3}, the concept of time in quantum theory remains elusive. This issue has been explored in several directions. The relation between time and energy, the physical quantity that is time-invariant in closed systems, has led to fundamental limitations in the form of quantum speed limits \cite{Pauli,MT45,ML98,Marvian2016,Pires2016}. Another approach has aimed to promote time from a mere classical parameter to a fully quantum description \cite{PW83,W84,GML,aki11,BW}. Notably, quantum evolution is here captured via the notion of correlations. Finally, various models of quantum systems designed to measure time, i.e.\ quantum clocks, have been proposed; see, for example, Refs.~\cite{Peres,motion,opticlocki,ralph}. These models typically consider a specific degree of freedom of a quantum system, prepared in a judiciously chosen initial state, then subjected to a unitary evolution, and finally measured. The result is interpreted as a time interval measurement, whose precision can be related to the properties of the clock (e.g. its dimension \cite{ralph}). However, the procedures of the state preparation and the measurement are usually not discussed explicitly. These models thus allow one to measure a time interval, e.g. for implementing a given unitary operation (by timing an interaction). This functionality is analogous to a stopwatch, but cannot be considered a complete model of a quantum clock.

\begin{figure*}[t]
\begin{center}
\includegraphics[width=0.8\linewidth]{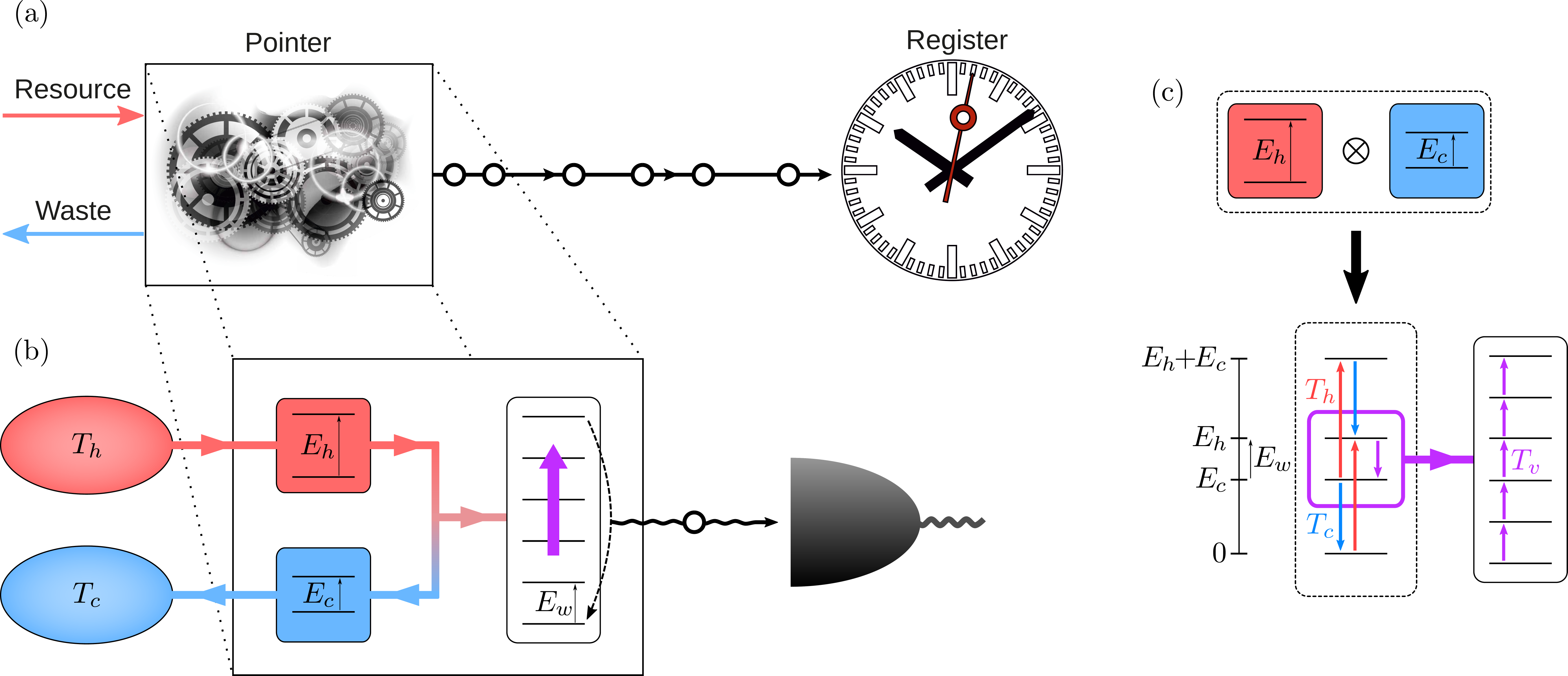}
\end{center}
\caption{(a)~A pointer system generates a time-ordered sequence of events that are recorded and displayed by the register. (b)~We consider a pointer comprising a two-qubit heat engine that drives a thermally isolated load up a ladder, whose highest-energy state undergoes radiative decay back to the ground state. Photons are thus repeatedly emitted and registered by a photodetector as ticks of the clock. (c) A virtual qubit is a pair of states in the engine's two-qubit Hilbert space whose energy splitting is resonant with the ladder. The thermal baths drive population into the virtual qubit's higher-energy state and out of its lower-energy one, creating a population inversion described by a negative virtual temperature. Hence, placing the virtual qubit in thermal contact with the ladder forces the load upwards, thereby performing work.}
\label{clockFig}
\end{figure*}

Indeed, a crucial feature of a clock (as opposed to a stopwatch) is to continuously provide a time reference to an external observer. It is thus essential that any complete model of a quantum clock explicitly specifies the process of information read-out. This leads us to consider a clock as a bipartite system~\cite{Hourglass,rank}, shown in Fig.~\ref{clockFig}(a). The first part of the clock is the \textit{pointer}, i.e.\ a subsystem whose internal dynamics are effectively dictated by the passage of time. The second part is the \textit{register}, which stores classical information obtained about the evolution of the pointer, thereby mediating the transfer of information from the system to an external observer. The pointer is designed to produce a sequence of signals, which are then recorded by the register as \textit{ticks}. 

It follows that there is an asymmetric flow of information between the two parts of the clock, which makes the process irreversible (and singles out a direction for the flow of time). This naturally connects the problem to the second law of thermodynamics \cite{Boltz}, because irreversibility is associated with the generation of entropy. One therefore expects that the suitability of a system for measuring time implies a corresponding propensity to produce entropy. However, a precise relationship between entropy production and clock performance has not yet been demonstrated.

In fact, we will show that such a relationship unavoidably becomes apparent when considering a more general question: what are the \textit{minimal} resources required to maintain a quantum clock? In order to answer this question, we {consider} an \textit{autonomous} quantum clock, i.e. a { self-contained} device working without any external control or timing. 
{The clock must be an isolated system evolving according to a time-independent Hamiltonian~\cite{rank}. Moreover, the resources powering the clock should not themselves require another clock to be prepared.}
Specifically, we {discuss} a natural class of autonomous clocks driven by minimal non-equilibrium resources, namely the flow of heat between two thermal reservoirs. 
{In particular, our model makes explicit the physical mechanism of the clock's operation, including its initialization and power supply.}
{We make use of} thermodynamical concepts in order to analyse the clock as an autonomous thermal machine \cite{linden10,LevKos12,virtual}, with the goal of producing a series of regular ticks.

{This approach allows us to} show that the clock's irreversible entropy production dictates fundamental limits on its performance. The performance of the clock is characterized by i) its \emph{resolution}, i.e.\ how frequently the clock ticks, and ii) its \emph{accuracy}, i.e. how many ticks the clock provides before its uncertainty becomes greater than the average time between ticks. We find that a given resolution and accuracy can be simultaneously achieved only if the rate of entropy production is sufficiently large; otherwise, a trade-off exists whereby the desired accuracy can only be attained by sacrificing some resolution, or vice versa. Furthermore, in the regime where the resolution is arbitrarily low, the accuracy is still bounded by the entropy production, suggesting a quantitative connection between entropy production and the clock's arrow of time. Note that here the relevant entropy production is not associated with measurements or erasure of the register, but rather with the evolution of the pointer system itself. In the following, we illustrate this behaviour by explicitly calculating the dynamics of a simple clock model. We then present a conjecture, backed up by general thermodynamic arguments, that such trade-offs are  exhibited by any implementation of an autonomous clock.

\section{Autonomous quantum clocks}
\label{sec:autonomous}

Our objective is to find the fundamental limits on quantum clocks. To that end, we {consider} \textit{autonomous} clocks, i.e.\ those which are complete and self-contained. In particular, the operation of the device should not require any time-dependent control that would necessitate another, external clock. This allows all resources needed for timekeeping to be carefully accounted for. In this section we discuss some of the general features of autonomous clocks, before specifying a particular model in Section \ref{sec:minimalModel}.

An autonomous clock evolves under a time-independent Hamiltonian, such that a steady stream of ticks are recorded at the register, as depicted in Fig.~\ref{clockFig}. The process by which information is transferred from pointer to register should be effectively irreversible, in order to ensure the unidirectional flow of time as recorded by the register. In addition, this process should occur spontaneously, i.e.\ without any external intervention or time-dependent coupling between the pointer and register. To ensure that the probability of this spontaneous process is larger than that of its time-reverse, the free energy of the pointer must decrease. Therefore, in order to continue producing ticks, the clock needs a source of free energy driving it out of equilibrium.

In principle, any nonequilibrium quantum system could provide the free energy needed to power a clock. However, a large class of nonequilibrium states are difficult to prepare in practice unless a clock is already available, e.g.\ so that a resonant driving field can be applied for a known period of time. We exclude such resource states in order to ensure fair bookkeeping, i.e.\ the resources' initial preparation should not itself require time measurements. It is also clearly desirable --- yet inessential --- that such resources be naturally abundant or otherwise easy to generate.

Here we argue that the \textit{minimal} nonequilibrium resource consists of two thermal reservoirs at different temperatures. Indeed, the presence of one heat bath is unavoidable, since this represents the environment at ambient temperature $T_c$. Furthermore, a second reservoir at temperature $T_h>T_c$ can be prepared deterministically without detailed understanding of the bath's internal structure and without any well-timed operations. This is because the thermal state represents a condition of minimal knowledge \cite{Jaynes} towards which generic quantum systems (i.e.\ those not integrable nor many-body localised) equilibrate \cite{gogolinio}.  In this sense the minimal out of equilibrium resource is an equilibrated (thermalized) resource with a higher average energy content than the environment. Any other potential resource for the clock would feature lower entropy at equal energies and thus additional knowledge/control to prepare. In the following, we base our quantitative analysis on clocks driven by thermal baths. However, we emphasize that the notion of an autonomous clock is more general and could be extended to various different scenarios and resource states.

\section{Minimal thermal clock model}
\label{sec:minimalModel}

We now specialize to a concrete model of an autonomous quantum clock where the pointer is driven by the heat flow between two thermal baths. For simplicity, we base our model on the smallest quantum heat engine that was introduced in Ref.~\cite{virtual} (see Appendix \ref{apptq} for a detailed description).

The machine consists of two qubits, each coupled to an independent thermal bath, as depicted in Fig.~\ref{clockFig}(b). The first qubit, connected to the hot bath at temperature $T_h$, has energy gap $E_h$. The second qubit is connected to a cold bath at temperature $T_c < T_h$ and has energy gap $E_c < E_h$. The engine delivers work to a load, represented by a system with $d$ equally spaced energy levels, i.e. a discrete ladder, with energy spacing $E_w = E_h-E_c$. 

The temperature difference between the two baths induces a heat current in the system from the hot qubit to the cold one. This flow of heat delivers energy to the load, causing it to ``climb'' the ladder. The action of the machine can be understood in terms of the resonant exchange of energy between the load and a \textit{virtual qubit}~\cite{virtual}. This virtual qubit is a special pair of states in the engine's Hilbert space that are coupled to the ladder, illustrated in Fig.~\ref{clockFig}(c). Assuming that the engine-ladder coupling is weak, the populations of the virtual qubit states are thermally distributed at the \textit{virtual temperature}
\be \label{virtualtemp}
k_B T_v = \frac{E_h - E_c}{ \beta_h E_h - \beta_c E_c},
\ee
where $\beta_{c,h} = 1/k_B T_{c,h}$. That is, the virtual qubit's states are occupied in the ratio $p_1/p_0 = {{\rm e}}^{-\beta_v E_w}$, where $p_1$ ($p_0$) denotes the population of the state with higher (lower) energy and $\beta_v = 1/k_B T_v$. Therefore, whenever the virtual qubit has a negative temperature, i.e. a population inversion, the load moves up the ladder as it ``thermalizes'' with the virtual qubit. The virtual temperature is conveniently parametrised by the virtual qubit's population bias
\be
\label{virtualbias}
Z_v  = \frac{p_0 - p_1}{p_0+p_1} = \tanh( \beta_v E_w /2 ),
\ee
which plays a central role in characterizing the performance of our clock, as we show below.

To complete the description of our clock, we must specify how the pointer interacts with the register. The top level of the ladder is assumed to be unstable, and decays to the ground state by emitting a photon at energy $E_\gamma = (d-1) E_w$. This photon is then detected at the register, which in turn makes the clock tick. Note that the presence of the decay channel also allows in principle for the reverse process. However, we assume that the background temperature satisfies $k_BT_c \ll E_\gamma$ so that such processes are negligible.

In summary, the flow of heat through the engine drives the load up the ladder, which eventually reaches the top level and decays back to the ground state while emitting a photon. The process is repeated, thus generating a steady stream of photons that are recorded by the register as ticks of the clock. Importantly, the evolution of the ladder's energy is probabilistic, leading to a stochastic sequence of ticks. The distribution of ticks depends in particular on the dimension of the ladder $d$ and the bias $Z_v$. Intuitively, if the bias is small ($Z_v$ negative but close to zero), the probability for the load to move up is only marginally larger than its probability of going down. The probability distribution over the levels of the ladder thus rapidly becomes quite broad, which makes the clock tick slowly and at irregular time intervals. On the other hand, if $Z_v \rightarrow -1$, i.e. the virtual qubit has essentially complete population inversion, then the probability for the ladder population to move downward is negligible, resulting in shorter and more regular time intervals between ticks.

\section{Performance of the clock}\label{sec:performance}

\begin{figure*}
\begin{minipage}{0.329\linewidth}
\flushleft {\scriptsize (a)} 

\includegraphics[width=\linewidth, trim = 30mm 90mm 30mm 90mm, clip]{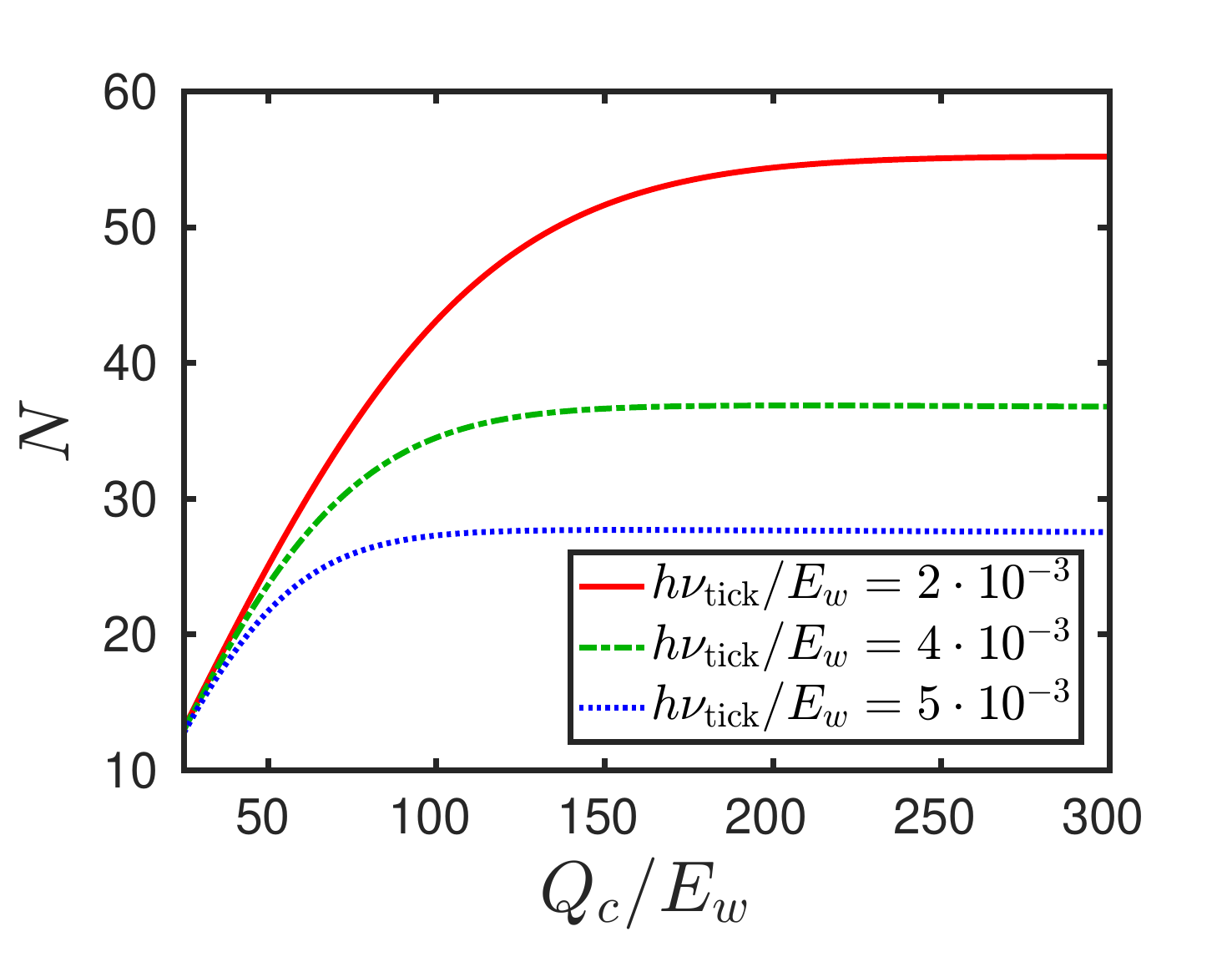}
\end{minipage}
\begin{minipage}{0.329\linewidth}
\flushleft {\scriptsize (b)} 
\includegraphics[width=\linewidth,  trim = 30mm 90mm 30mm 90mm, clip]{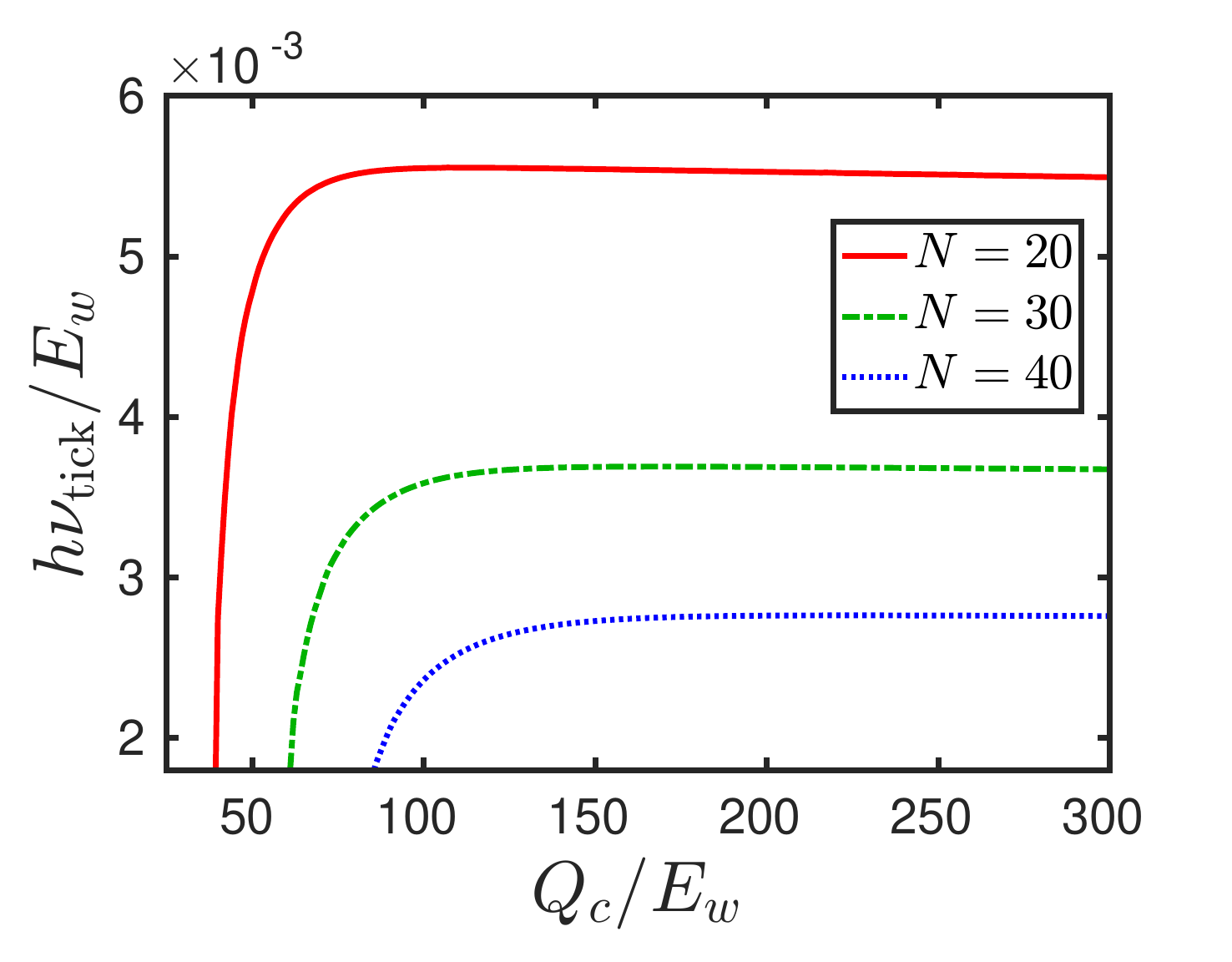}\end{minipage}
\begin{minipage}{0.329\linewidth}
\flushleft {\scriptsize (c)} %\vspace{-5mm}

\includegraphics[width=\linewidth,  trim = 30mm 90mm 30mm 90mm, clip]{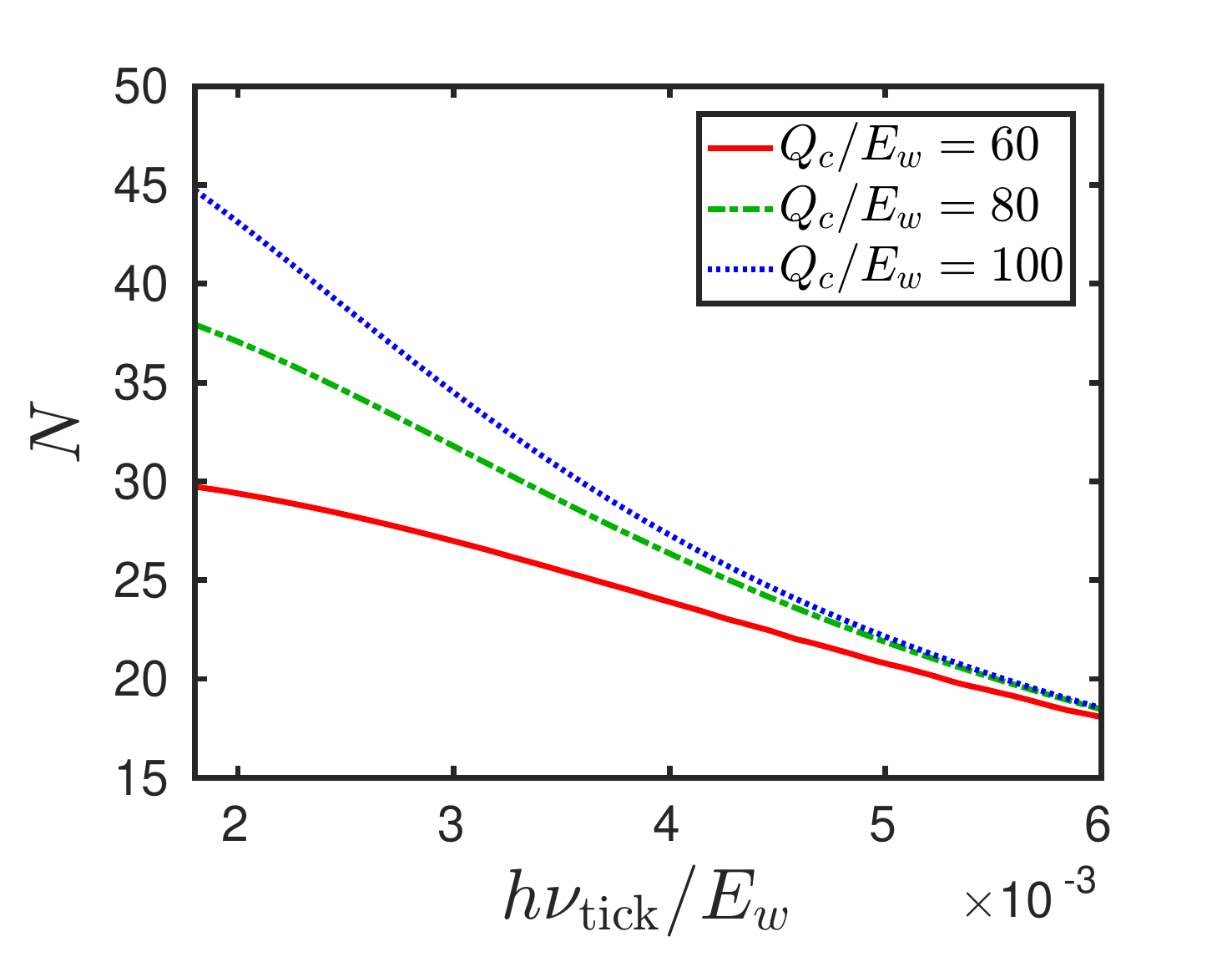}
\end{minipage}
\caption{Illustration of the fundamental trade-off between the dissipated heat and the achievable accuracy and resolution. (a)~Accuracy $N$ as a function of dissipated heat per tick $Q_c$, for various values of the resolution $\nu_\mathrm{tick}$. 
At low energy, the accuracy increases linearly with the dissipated energy, independently of the resolution. However, for higher energies, the accuracy saturates.
(b)~Resolution $\nu_\mathrm{tick}$ as a function of dissipated heat per tick $Q_c$, for various values of the accuracy $N$. The resolution first increases with dissipated energy, but then quickly saturates to a maximal value. 
(c)~Trade-off between accuracy and resolution when the energy dissipation rate is fixed.
The data are computed for fixed values of $k_BT_c = E_w$, $k_BT_h= 1000E_w$ and $g = \hbar\gamma = \hbar\Gamma = 0.05E_w$, while the ladder dimension $d$ and cold qubit energy $E_c$ are varied independently. Note that $d \geq 10$ for all of the plotted points, thus $k_{B} T_c = E_w \ll E_\gamma = (d-1) E_w$ and we can safely ignore the absorption of a photon (i.e. the reverse of the decay process).
\label{resolutionPlot}}
\end{figure*}

In order for the clock to deliver ticks, the engine must raise the ladder's energy and necessarily dissipate energy into the cold bath. Our goal now is to relate the performance of the clock to this dissipated energy, which is closely related to the entropy production. Specifically, we consider here the heat dissipated into the cold bath per tick of the clock
\begin{equation}
\label{Qc}
Q_c = (d-1) E_c.
\end{equation}
Note that this quantity, rather than the heat supplied to the machine per tick [$Q_h = (d-1)E_h$], represents the fundamental minimum energy expenditure associated with one tick of the clock. This is because, in principle, a large part of the energy $E_\gamma$ carried away by the emitted photon could be captured and recycled (e.g.\ dumped back into the hot bath). Consequently, the dissipated heat~\eqref{Qc} is associated with an irreversible entropy production of at least $\beta_cQ_c$ per tick.

The performance of our autonomous clock is quantified by the \textit{resolution} and \textit{accuracy} of its ticks. By \textit{resolution}, we refer to the average number of ticks the clock provides per unit time. The ticks are not distributed regularly, and we characterize the \textit{accuracy} by the number of ticks provided before the next tick is uncertain by the average time interval between ticks~\cite{footrank}.

For our model of the autonomous clock, we assume that after each spontaneous emission event, the entire pointer is reset to its initial state --- specifically, a product state with the ladder in its ground state and the engine qubits in equilibrium with their respective baths. This approximation is valid in the weak-coupling limit, where the engine qubits are minimally perturbed by their interaction with the ladder. The ticks of the clock can therefore be described as a renewal process, i.e.\ the time between any pair of consecutive ticks is statistically independent from, and identically distributed to, the time between any other pair of consecutive ticks. 

Now, let the distribution of waiting times between two consecutive ticks be characterized by the mean $t_{\rm tick}$ and the standard deviation $\Delta t_\mathrm{\rm tick}$. The resolution of the clock is then
\be 
\label{resolutionDef}
\nu_\mathrm{tick} = 1/t_{\rm tick},
\ee 
i.e.\ the average number of ticks the clock provides per second. The accuracy is the number of ticks $N$ such that the uncertainty (standard deviation) of the $N^{\mathrm{th}}$ tick time is equal to the average time between ticks. Since the waiting times are independent, the uncertainty in the time of the $n^{\rm th}$ tick is simply $\sqrt{n}\Delta t_\mathrm{tick}$, and therefore
\be 
\label{accuracyDef}
N = \left(\frac{t_{\rm tick}}{\Delta t_{\rm tick}}\right)^2.
\ee

Fig.~\ref{resolutionPlot} illustrates the intimate relationship between the accuracy $N$ and the resolution $\nu_\mathrm{tick}$ versus the dissipated energy $Q_c$, calculated by numerical solution of the equations of motion (see Appendix~\ref{appdyn}). 
We find that, for a given amount of dissipated energy, there is a trade-off between accuracy and resolution. In other words, engineering a good clock featuring both high accuracy and high resolution requires a large amount of energy to be dissipated and thus a higher production of entropy per tick. This is nicely illustrated in Fig.~\ref{resolutionPlot} c), which showcases the nature of entropy production as a resource. The curves for different entropies are clearly ordered, i.e. more entropy implies either more resolution or more accuracy can be achieved. It is interesting to note, however, that the relationship between the two is non-trivial and the trade-off features non-linear dependencies.

Finally, we note that in the regime of low energy dissipation the relationship between accuracy and entropy production at fixed resolution is directly proportional as seen in Fig.~\ref{resolutionPlot} a). In the next section we will recover this behaviour analytically in the weak coupling regime.

\section{Accuracy in the weak-coupling limit}

We now investigate the relationship between accuracy and dissipated power by an alternate approximate analysis, valid when the interaction between the engine and the ladder is weak. In this regime, the accuracy is limited by the dissipated power and \emph{the dimension} of the ladder, while the resolution is not focused upon. This is in contrast to Fig.~\ref{resolutionPlot}(a), where the resolution is fixed, and the dimension is allowed to vary. In particular, we show that the accuracy is essentially independent of the details of the clock's dynamics, being determined only by the bias of the virtual qubit $Z_v$ and the ladder dimension $d$.

Focusing on the ladder, its evolution can be approximated by a biased random walk, induced by the interaction with the virtual qubit. This is easily understood through the fact that the resonant interaction with the virtual qubit cannot induce any coherence on the ladder. Moreover, the resonance is exactly at the energy of a transition of one step up or down, and independent of the ladder's position. The rates at which the ladder population moves upwards ($p_{\uparrow}$) or downwards ($p_{\downarrow}$) satisfy $p_{\uparrow}/p_{\downarrow} = {{\rm e}}^{-\beta_v E_w}$ as a consequence of detailed balance. This description of the clock is derived in Appendix \ref{apprwa} as a perturbative approximation to the two-qubit engine, which becomes exact in the limit of vanishingly small engine-ladder coupling. We also make the simplifying assumptions that the clock ticks as soon as the load reaches the top of the ladder, and that $d$ is large enough for reflections from the boundaries of the ladder to be negligible.

Under the foregoing approximations, the resolution is given by 
\begin{equation}
\label{stochasticResolution}
\nu_{\rm tick} = \frac{p_{\uparrow} - p_\downarrow}{d}.
\end{equation}
Quite intuitively, the resolution is inversely proportional to the dimension $d$, corresponding to the ``height'' of the ladder, but is proportional to the difference of transition rates $p_{\uparrow}-p_{\downarrow}$, which quantifies the ``speed" at which the load climbs. 

On the other hand, as demonstrated in Appendix~\ref{apprwa}, the accuracy is given by
\begin{equation}\label{accuracyprelude}
	N = d |Z_v|,
\end{equation}
which is entirely independent from the clock's overall dynamical time scale, set by the rates $p_{\uparrow,\downarrow}$. Instead, the accuracy depends only on the dimensionless quantities $Z_v$ and $d$. In turn, the bias $Z_v$ encapsulates the dependence of the clock's accuracy on the dissipated heat. In the case of our model, using equations \eqref{virtualtemp} and \eqref{virtualbias}, the accuracy is given by
\begin{equation}
\label{accuracypowerheuristic}
N = d \tanh \left[ \frac{(\beta_c- \beta_h) Q_c - \beta_h E_{\gamma}}{2d} \right].
\end{equation}
Note, however, that the relation between $Z_v$ and the heat exchanged with the two baths is more general than the model considered here \cite{Silva16} (see Appendix \ref{appg} for a discussion). It follows that accuracy in the weak-coupling limit depends on the amount of dissipated heat but not on the dissipation rates. 

\begin{figure}
\includegraphics[width=0.8\linewidth,  trim = 30mm 90mm 30mm 90mm, clip]{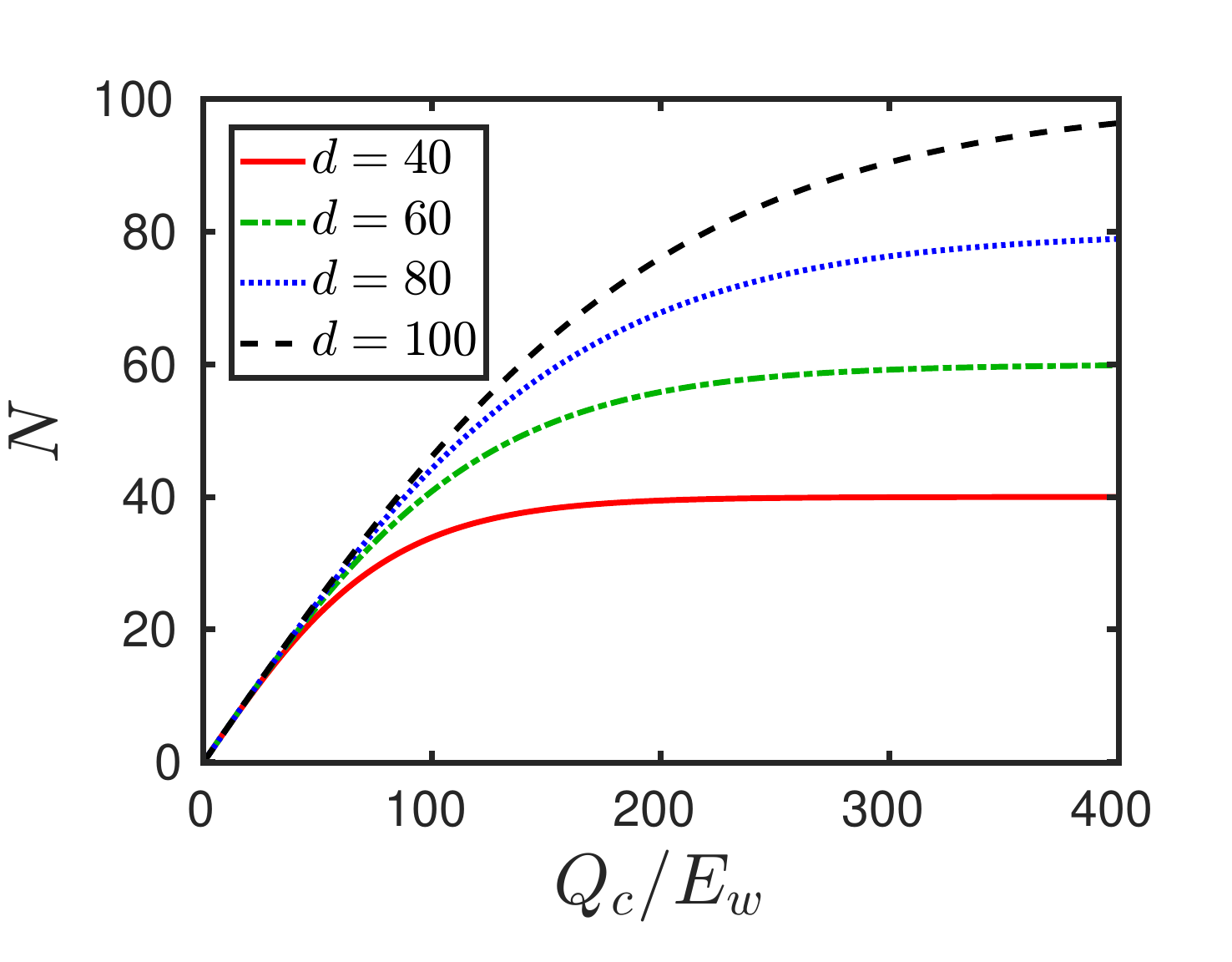} 
\caption{Accuracy $N$ versus dissipated energy $Q_c$ for various values of the dimension $d$ of the ladder, according to the approximation (\ref{accuracypowerheuristic}) with the same bath temperatures as in Fig.~\ref{resolutionPlot}. \label{dimension}} 
\end{figure}

The behaviour described by equation~\eqref{accuracypowerheuristic} is illustrated in Fig.~\ref{dimension}, where we plot the accuracy versus the dissipated energy for fixed dimension. We observe that the accuracy first increases linearly but eventually saturates to its maximum value $N = d$. Indeed, increasing $Q_c$ leads to a stronger bias in the virtual qubit, saturating at $|Z_v| \rightarrow 1$ as $Q_c \rightarrow \infty$. Thus the accuracy is limited by both the dimension $d$ and the dissipated energy $Q_c$. Hence, achieving a certain accuracy requires a minimum dimension as well as a minimum dissipated energy per tick.

Even if the dimension is unbounded, we find that the dissipated energy still imposes a fundamental limitation. Taking the limit $d\rightarrow\infty$, the accuracy is linearly dependent on the dissipated heat:
\begin{equation}
	N \rightarrow \frac{(\beta_c- \beta_h) Q_c - \beta_h E_{\gamma}}{2}.
\end{equation}

Noting that $Q_h = Q_c + E_\gamma$, we can recast the above in the illustrative form
\begin{equation}\label{arrowoftime}
	N \rightarrow \frac{\beta_c Q_c - \beta_h Q_h}{2} = \frac{\Delta S_{tick}}{2},
\end{equation}
where $\Delta S_{tick}$ is the increase in the entropy of the clock in a single tick. We may interpret the regularity of each tick as representative of the strength of the arrow of time.  Thus \eqref{arrowoftime} quantifies in a concrete manner the connection between the arrow of time of a clock and its \textit{irreversibility}.

\section{Fundamental limits of general autonomous clocks}

The simple thermal clock model we discuss above illustrates the fact that our ability to accurately and precisely measure time necessarily generates an increase of entropy (via heat dissipation). Equivalently, this implies an intrinsic work cost for measuring time. It is natural to ask whether the connection between clock performance and entropy production is a specific aspect of our model, or on the contrary a universal feature of any procedure for measuring time. Below, we argue in favor of the latter: any autonomous clock must increase entropy.

The core insight underlying our argument is that, as discussed in Section~\ref{sec:autonomous}, the ticks of any autonomous clock involve a spontaneous and effectively irreversible transition in a pointer system, thus inducing a corresponding change in the register to which it is coupled. 
In order to bias the forward transition in favour of its time-reverse (i.e.\ to avoid the clock ticking ``backwards''), the transition must reduce the free energy of the pointer. Hence, for the clock to run continuously it needs access to a system out of thermal equilibrium that can replenish the free energy of the pointer. Now, the essential question is whether it is possible for the clock to convert this free energy into ticks with perfect efficiency, i.e.\ without increasing entropy.

Let us first discuss this question in the context of clocks driven by thermal baths. It is clear that beyond the specific model we have studied, one could consider more general designs for the thermal machine. The basic necessary ingredient is simply the ability to move the population of the pointer out of equilibrium, so that an unstable level generates a tick. This transition is biased in the forward direction so long as the unstable level is much higher in energy than the thermal background. Such a mechanism can indeed work for a variety of physical implementations of the pointer (i.e.\ with a more complex level structure). The ladder could comprise multiple levels which trigger a decay, while the machine could feature more than two qubits. 

Nonetheless, all these possible extensions and more sophisticated designs will still have to comply with the basic laws of thermodynamics. In particular, the efficiency of the conversion of energy to a tick is fundamentally bounded by the Carnot efficiency $\eta_C=1-T_c/T_h$. Moreover, this maximal efficiency can only be achieved in a limit where the power vanishes, corresponding to the regime where the machine works reversibly. A finite power, however, is essential for the resolution of any autonomous clock: a clock working at Carnot efficiency ticks infinitely slowly. Hence, even in the rather artificial regimes of $T_c\rightarrow0$ or $T_h\rightarrow\infty$, the requirement of a finite resolution implies a minimal dissipated heat, and thus a minimal entropy production. 

It is also possible to consider more general nonequilibrium resources to power the clock. In order to satisfy the requirement of autonomy, such resources should not themselves need any well-timed control in order to be produced. In principle, it is conceivable that such a resource could allow the clock to achieve higher efficiency than is possible with thermal driving. However, an autonomous clock that does not generate any entropy but nonetheless has finite resolution would constitute an autonomous machine operating at finite power with unit efficiency. Therefore, if the performance of autonomous quantum clocks is not always associated with a fundamental entropy production, then the prospect of quantum machines is far more revolutionary than is widely believed at present.

Finally, it is also worth pointing out that, while we focus here on a specific source for the entropy production of the clock (namely the heat dissipated by the thermal machine driving the clock), there will be generally additional energy costs required for operating the clock. In particular, the preparation (and reset) of the initial state of the register will generate entropy due to Landauer's erasure principle \cite{Land,QLand}. 

Even if the qualitative bound \eqref{arrowoftime} derived in our work represents a fundamental limit for any clock, it still underestimates the necessary costs of running the best clocks available today. For instance, a typical atomic clock\cite{atomic} runs at resolutions of the order of $10^{10}$ Hz, and an accuracy of $10^{16}$ seconds before being off by a second. \eqref{arrowoftime} would imply imply a minimal power consumption for such a clock of the order of about $50 \mu W$. In practice, the real costs are orders of magnitude higher. This is similar to the case of information erasure: even though Landauer's principle is the only known fundamental limit, current erasure techniques operate far less efficiently.

\section{Conclusion and outlook}

Our work represents a first step towards rigorously characterizing the necessary resources and limitations of the process of timekeeping. 
In a nutshell, we introduced the concept of autonomous quantum clocks to discuss these questions, and argued that the measurement of time inevitably leads to entropy increase. Moreover, we discussed explicitly a simple model of an autonomous quantum clock, and found that the amount of entropy produced represents an actual resource for measuring time. Every unit of heat dissipated can be spent to increase either the accuracy or the resolution of the clock. Additionally, the dimension of a key constituent of the clock (the ladder) imposes a limit on the achievable accuracy and resolution, independently of the amount of dissipated heat. 
In other words, in analogy to the findings of \cite{rank,ralph}, the Hilbert space dimension imposes a fundamental constraint on the performance of the clock. Reaching this optimal regime requires a minimal rate of entropy production. This provides a quantitative basis for the intuitive connection between the second law of thermodynamics and the arrow of time (see, for example, Refs.~\cite{B,L}). In order to measure how much time has passed, we inevitably need to increase the entropy of the universe from the perspective of the register. 

These considerations only concern here the scenario of minimal autonomous clocks, i.e. where the resources exploited to operate the clock are simply two thermal baths at different temperatures. While these arguably represent the most abundant resources found in nature \cite{gogolinio}, it would be interesting to consider other quantum systems, e.g.\ with multiple conserved quantities \cite{yelena,oppenheim,rudolph,jorg}.
More broadly, the relevant question is to what extent our choice of free resources impacts our ability to measure time. For instance, one could consider more general passive states \cite{pusz}, that would commute with the system Hamiltonian and thus satisfy the requirement of autonomy. 
Thermal clock models can furthermore be used to work out the thermodynamic cost of controlling other quantum systems \cite{ralph,Campbell,kammi} in an autonomous fashion, i.e.\ implementing locally apparent time-dependent Hamiltonians by coupling to an autonomous thermal clock.
Moreover, operating two clocks in parallel could lead to a drastic enhancement of the clock's performance. While classical clocks running in parallel would not offer any fundamental improvement, one could consider quantum resources that feature coherence or entanglement \cite{entclock,network}. Could these genuine quantum phenomena be used to increase our ability to measure time? We look forward to future research in this direction.

\textbf{Acknowledgements.} We are grateful to \"Amin Baumeler, Nicolas Gisin, Patrick Hofer, Daniel Patel, Sandu Popescu,  Gilles P\"utz, Sandra Rankovic Stupar, Renato Renner, Christian Klumpp and Stefan Wolf for fruitful discussions. MH acknowledges funding from the Swiss National Science Foundation (AMBIZIONE PZ00P2$\_$161351) and the Austrian Science Fund (FWF) through the START project Y879-N27. MW and MTM acknowledge funding from the UK research council EPSRC. RS and NB acknowledge the Swiss National Science Foundation (Starting grant DIAQ, grant 200021$\_$169002, and QSIT). PE acknowledges funding by the European Commission (STREP RAQUEL), the Spanish MINECO, projects FIS2008-01236 and FIS2013-40627-P, with the support of FEDER funds, the Generalitat de Catalunya CIRIT, project 2014-SGR-966, the Swiss National Science Foundation (SNF) through the project `Information and Physics' and the National Centres of Competence in Research Quantum Science and Technology (QSIT).

\appendix

\section{Description of the two-qubit heat engine}	\label{apptq}
	
Here we give a detailed description of the two-qubit heat engine of Ref. \cite{virtual}, which represents the pointer of the autonomous quantum clock. The machine consists of two qubits, each one connected to a thermal bath. The first qubit with energy gap $E_h$ is connected to the bath at $T_h$. The second qubit is connected to the bath at $T_c$ and has energy gap $E_c < E_h$. The engine is connected to a $d$-dimensional ladder, featuring equally spaced energy levels (with spacing $E_w$), which is not connected to any heat bath. The free Hamiltonian of the total system (two qubits and ladder) is thus given by
\be 
H_0 = \sum_{j =h,c} E_j \ket{1}_j \bra{1}  +  \sum_{k=0}^{d-1} k E_w \ket{k}_w \bra{k},
\ee
where $\ket{1}_j$ denotes the excited state of qubit $j=h,c$, and $\ket{k}_w$ denotes the state of $k$-th level of the ladder. As a design constraint we take that 
\be \label{design}
E_h = E_c + E_w.
\ee
Hence the following energy levels of the total system are degenerate in energy: $\ket{0}_c \ket{1}_h \ket{k}_w$ and $\ket{1}_c \ket{0}_h \ket{k+1}_w$. This allows for energy to be exchanged between the qubits and the ladder. Specifically, we consider the interaction Hamiltonian
\be \label{Hint}
H_{\text{int}} = g \sum_{k=0}^{d-1}  \,  ( \ket{1}_c \ket{0}_h \ket{k+1}_w     \bra{0}_c \bra{1}_h \bra{k}_w  + \mathrm{h.c.} ).
\ee
The machine will be operated in the weak coupling regime, i.e. $g \ll E_c, E_w$. Note that our design constraint on the energies \eqref{design} ensures that $H_{\text{int}}$ has a significant effect even in the weak-coupling regime. Henceforth we refer to the joint system of ladder and engine as the pointer, since it will be the system from which the register will derive information reflecting the passage of time.

The functioning of the engine can be understood intuitively as follows. The temperature difference between the baths induces a heat flow from the first qubit (at $T_h$) to the second (at $T_c$). This heat flow is made possible by our design constraint \eqref{design}. Specifically, a quantum of energy $E_h$ from the first qubit can be transferred to a quantum of energy $E_c$ in the second qubit, while the remaining energy $E_h - E_c = E_w$ is transferred to the ladder. This process corresponds to the first term in the interaction Hamiltonian \eqref{Hint}. Indeed the reverse process is also possible, represented by the second term in \eqref{Hint}. For the engine to deliver work (i.e. to raise the energy of the ladder), we need to ensure that the first process is more likely than the second. This can be done by judiciously choosing the parameters (energies and temperatures) as we will see now.

We follow the approach of Ref. \cite{virtual}, which captures in simple and intuitive terms the effect of the two-qubit engine on the ladder \cite{footnote1}. In order to bias the transition in the direction
\be \label{transition}
    \ket{0}_c \ket{1}_h \ket{k}_w   \rightarrow  \ket{1}_c \ket{0}_h \ket{k+1}_w ,
\ee
we simply demand that the probability $p_1$ of occupying the state $\ket{0}_c \ket{1}_h$ is larger than the probability $p_0$ of occupying the state $\ket{1}_c \ket{0}_h$; recall that the ladder is only weakly connected to the ambient heat bath. As the machine works in the weak-coupling regime, these probabilities basically depend only on the baths' temperatures and the qubits' energies, the state of each qubit being close to a thermal state at the temperature of the corresponding bath. Hence, the transition \eqref{transition} is biased assuming that
\be  \label{condition}
\frac{E_h}{T_h} < \frac{E_c}{T_c}
\ee
The effect of the engine on the ladder is determined by the two states $\ket{0}_c \ket{1}_h$ and $\ket{1}_c \ket{0}_h$, which define the machine's virtual qubit. The engine simply places the load in thermal contact with the virtual qubit, which has energy gap $E_h-E_c = E_w$, hence resonant with the ladder's energy spacing, and virtual temperature determined by the population ratio $p_1/p_0 = {\rm e}^{-\beta_v E_w}$. The load will thus effectively ``thermalize'' with the virtual qubit. This causes the load to climb the ladder so long as the bias \eqref{virtualbias}, or equivalently the virtual temperature \eqref{virtualtemp}, is negative. Indeed, one can immediately check that the condition \eqref{condition} is satisfied whenever the virtual qubit has a negative bias.

\section{Dynamics of the clock}\label{appdyn}

In order to model the dynamics of the pointer and com- pute the distribution of ticks, we use the following master equation formulation.
The effect of each reservoir on its corresponding qubit is represented by the superoperator
	\begin{equation}\label{QubitDissLj}
	\LL_j = \gamma_j \DD[\sigma_j] + \gamma_j \e^{-\beta_j E_j}\mathcal{D}[\sigma_j^\dag],
	\end{equation}
	for $j = h,c$. Here we defined the qubit lowering operators $\sigma_j = \ket{0}_j\bra{1}$, and the dissipator in Lindblad form
	\begin{equation}\label{generalDissipator}
	\DD[L] \rho = L \rho L^\dag - \frac{1}{2} \left\{ L^\dag L,\rho \right\}.
	\end{equation}
The rates $\gamma_{h,c}$ determine the overall time scale of the dissipative processes acting on the two engine qubits.
	
In addition, the ladder system couples to a reservoir of electromagnetic field modes at temperature $T_c$. The ladder is designed so that only the highest energy transition $\ket{d-1}_w\to \ket{0}_w$ couples significantly to the electromagnetic field. This transition is associated with the emission of a photon having energy $(d-1) E_w$, while $\Gamma$ is the spontaneous emission rate. A photo-detector registers the emitted photon, producing a macroscopically measurable ``tick". The detector is assumed to work with perfect efficiency and negligible time delay. Furthermore, the background temperature $T_c$ is assumed to be low enough that we can ignore the reverse transition $ \ket{0}_w \to \ket{d-1}_w$, wherein the ladder absorbs a photon while in the ground state, i.e.\ we require that $k_B T_c\ll (d-1) E_w$.
	
To quantify the ticks of the clock, in principle one would have to keep track of the density operator of the pointer $\rho(t)$ for all times $t$. However, as argued in the main text, in the weak-coupling regime, the qubit states do not change appreciably from the thermal states corresponding to equilibrium with their respective reservoirs. Each tick is therefore independent of the previous ticks, and one can study the relevant quantifiers of the clock (i.e. resolution and accuracy) from the probability distribution in time of a single tick. 

We describe the dynamics of the clock in the ``no-click" subspace, i.e. the subensemble $\rho_0(t)$ conditioned on no spontaneous emission having occurred up to time $t$. We assume the pointer begins in the normalized state
	\begin{equation}\label{rho0Initial}
	\rho_0(0) = \frac{\e^{-\beta_h E_h \sigma^\dag_h \sigma_h}}{\mathcal{Z}_h} \otimes \frac{ \e^{-\beta_c E_c \sigma^\dag_c \sigma_c}}{\mathcal{Z}_c} \otimes \ket{0}_w \!\bra{0},
	\end{equation}
	where $\mathcal{Z}_{c,h}$ are the partition functions necessary for normalization. Equation~\eqref{rho0Initial} describes the situation where the qubits are in equilibrium with their respective reservoirs, and the ladder has just decayed and been reset into the ground state (i.e.\ the register has just ticked). The subsequent evolution of the conditional density operator $\rho_0(t)$ follows from the master equation ($\hbar = 1$)
	\begin{equation}\label{conditionalMasterEquation}
	\dt{\rho_0} = \ii \left ( \rho_0 H_\mathrm{eff}^\dag - H_\mathrm{eff}\rho_0 \right ) + \LL_h \rho_0 + \LL_c \rho_0,
	\end{equation}
	where the effective non-Hermitian Hamiltonian is given by $H_{\mathrm{eff}} = H_0 + H_{\mathrm{int}} + H_\mathrm{se}$, with spontaneous emission described by the contribution
	\begin{equation}
	\label{Heff}
	H_{\rm{se}} = - \frac{\ii \Gamma}{2} \ket{d-1}_w \!\bra{d-1}.
	\end{equation}
	
As a result of the non-Hermitian contribution, $\rho_0(t)$ does not stay normalized. The trace of the conditional density operator $P_0(t) = \Tr[\rho_0(t)]$ corresponds to the probability that a tick has not yet occurred. The probability density $W(t)$ of the waiting time between two consecutive ticks then follows from
	\begin{equation}
	W(t) = - \dt{P_0}.
	\end{equation}
For our purposes, we need only the mean and variance of the waiting time, which are given by
\begin{align}
\label{TtickDef}
t_\mathrm{tick} &= \int_{0}^\infty\mathrm{d}\tau\; \tau W(\tau),  \\
\label{DeltaTtick}
(\Delta t_\mathrm{tick})^2 &= \int_{0}^\infty\mathrm{d}\tau\; (\tau - t_\mathrm{tick})^2 W(\tau).
\end{align}

\section{Biased random walk approximation}\label{apprwa}

In this appendix we determine the accuracy of the autonomous clock from a stochastic model of the pointer's evolution. Specifically, we make two simplifying assumptions. Firstly, the evolution of the pointer is simplified to a continuous biased random walk of the ladder, with rates controlled by the populations of the virtual qubit of the two-qubit engine. That is, the ladder has a rate per unit time to move upward and a rate to move down, and the ratio of the rates is given by the ratio of populations of the virtual qubit. This is an accurate description in the regime where the thermal couplings are much larger than the interaction between the engine and the ladder and the spontaneous emission rate (see the following section for details). Under this assumption, the density operator of the ladder is diagonal, and can be replaced by a vector of populations of the energy levels. The second assumption is that the dimension of the ladder is large enough so that, for most of its evolution, the population distribution does not feel the boundedness of the ladder Hamiltonian. 

From the preceding arguments, the state of the ladder can described by a time-dependent probability distribution on a grid of integers (that label the energy levels) $q(n,t)$, where $n \in \mathbb{Z}$, $q(n,t)>0$, and $\sum_n q(n,t) = 1 $. The evolution is determined by the forward rate $p_\uparrow$ per unit time of jumping to the next integer, together with the backward rate $p_\downarrow$ of jumping to the previous integer. An equation of motion of the distribution can thus be constructed:
	\begin{align}
\dt{q(n,t)} = &\; p_\uparrow \; q(n-1,t) + p_\downarrow \; q(n+1,t) \notag \\ & - \, (p_\uparrow+p_\downarrow) \; q(n,t). \label{eqmotionstoch}
	\end{align}
	
In order to characterize the resolution and accuracy, we must understand how quickly the position of the ladder moves up, as well as how much it spreads on the way. We denote the mean and variance of the distribution by $\mu$ and $\sigma^2$ respectively 
	\begin{align}
	\label{muDef}
		\mu(t) &= \sum_n n \; q(n,t),\\
\label{sigDef}		
		\sigma^2(t) &= \sum_n \left( n-\mu(t) \right)^2 q(n,t).
	\end{align}
The speed of the ladder is determined by a simple calculation
	\begin{align}
		\dt{\mu(t)} &= \sum_n n \; \dt{q(n,t)} \notag \\
		& = p_\uparrow - p_\downarrow.
	\end{align}
	The variance may be similarly calculated from
	\begin{align}
		\dt{ \sigma^2(t)} = \sum_n & \left( \left( n - \mu(t) \right)^2 \dt{q(n,t)} \right . \notag \\ & \; -  \left . 2 \left( n - \mu(t) \right) \dt{\mu(t)} q(n,t)  \right).
	\end{align}
Using equations~\eqref{muDef} and \eqref{sigDef}, the second term can be shown to vanish, while the first term simplifies to
	\begin{equation}
	\dt{\sigma^2(t)} = p_\uparrow + p_\downarrow.
	\end{equation}
	
We are now in a position to find the relevant quantifiers of the clock. The average time between ticks is taken to be the time for the ladder to travel from the bottom to the top of its spectrum of $d$ eigenvalues,
	\begin{equation}
	t_{\rm tick} = \frac{d}{ \dd \mu(t) / \dd t} = \frac{d}{p_\uparrow - p_\downarrow},
	\end{equation}
	where for simplicity we replace $d-1$ by $d$, since the dimension of the ladder has been assumed to be large. The resolution $\nu_{\rm tick}$, i.e. the number of ticks per unit time, is the inverse of $t_{\rm tick}$,
	\begin{equation}
	\nu_{\rm tick} = \frac{p_\uparrow - p_\downarrow}{d},
	\end{equation}
	corresponding to equation~\ref{stochasticResolution}.
	
In the time taken for a single tick, the variance of the ladder will have increased by
	\begin{equation}
	\Delta \sigma^2 = t_{\rm tick} \frac{\dd \sigma^2(t)}{\dd t} = d \left( \frac{p_\uparrow + p_\downarrow}{p_\uparrow - p_\downarrow} \right).
	\end{equation}
Assuming the decay mechanism is good enough that the uncertainty in a single tick is determined solely by the uncertainty in when the ladder reaches the top (i.e. the variance), then the uncertainty in the time interval between consecutive ticks is simply
	\begin{equation}
	\Delta t_{\rm tick} = \frac{\sigma\left( t= t_{\rm tick} \right)}{\dd \mu(t)/\dd t} = \frac{\sqrt{d}}{p_\uparrow - p_\downarrow} \sqrt{ \frac{p_\uparrow + p_\downarrow}{p_\uparrow - p_\downarrow} }.
	\end{equation}
	The accuracy $N$ is defined as the number of ticks until the clock is uncertain by a single tick. This implies that the variance of the load has grown to the size of the entire ladder, $\sigma^2 = d^2$. It follows that
	\begin{align}
	N = d \left( \frac{p_\uparrow - p_\downarrow}{p_\uparrow + p_\downarrow} \right),
	\end{align}
which is equivalent to equation~\eqref{accuracyprelude} since $p_{\uparrow}/p_{\downarrow}= p_1/p_0$.
	
\section{Derivation of the biased random walk model}\label{apprw}
	
	Treating the pointer as a stochastic system is motivated by our understanding that the core of the machinery lies in the coupling of the ladder to the engine's virtual qubit, whose main effect is to create a bias such that the ladder's energy is more likely to increase than decrease. In this section, we place this (essentially classical) description of the pointer on a firmer footing, deriving it from the two-qubit engine model detailed above, working in the regime where the engine-ladder coupling $g$ and the spontaneous emission rate $\Gamma$ are both small in comparison to the thermal dissipation rates $\gamma_{c,h}$.
	
	In the limit of $\gamma_j \gg g,\Gamma$, we use the Nakajima-Zwanzig projection operator technique to derive an evolution equation for the conditional reduced density operator of the ladder $\rho_w(t) = \Tr_{h,c}[\rho_0(t)]$. We introduce the projector
	\begin{equation}
	\PP \rho_0(t) = \rho_w(t) \otimes \rho_h\otimes \rho_c,
	\end{equation}
	where $\rho_{h,c}$ denotes a local thermal state of the hot or cold qubit, for $j = h,c$, 
	\begin{equation}
	\label{localThermalStates}
	\rho_j = \frac{1}{\mathcal{Z}_j}\e^{-\beta_j E_j \sigma^\dag_j \sigma_j},
	\end{equation}
	while $\mathcal{Z}_j = 1 + \e^{-\beta_j E_j}$ is the corresponding partition function. Writing equation~(23) as $\dd\rho_0/\dd t = \LL\rho_0$, we decompose the Liouvillian as $\LL = \LL_0 + \HH_{\mathrm{se}} + \HH_{\mathrm{int}}$, where we defined the Hamiltonian superoperator
	\begin{equation}
	\label{HspontSuper}
	\HH_{\mathrm{se}} \rho = \ii\left (\rho H_\mathrm{se}^\dag - H_\mathrm{se}\rho\right ),
	\end{equation}
	and similarly for $ \HH_{\mathrm{int}}$. 
	
We transform the density operator to a dissipative interaction picture defined by $\tilde{\rho}_0(t) = \e^{-\LL_0 t} \rho_0(t)$. The time dependence of superoperators is given in this picture by $\tilde{\HH}_\mathrm{int}(t) = \e^{-\LL_0 t}\HH_\mathrm{int}\e^{\LL_0t}$ and $\tilde{\HH}_\mathrm{se}(t) =\HH_\mathrm{se}$. Following the standard perturbative argument \cite{BreuerPetruccione}, we obtain 
	\begin{equation}
	\label{RedFieldEquation}
	\dt{\PP \tilde{\rho}_0} = \HH_{\mathrm{se}} \PP  \tilde{\rho}_0(t) + \int_0^t \dd t' \; \PP \tilde{\HH}_\mathrm{int}(t) \tilde{\HH}_\mathrm{int}(t')\PP \tilde{\rho}_0(t'),
	\end{equation} 
	valid to second order in the small quantities $g$ and $\Gamma$. We now apply the Born-Markov approximation to the $t'$ integral above, extending the lower integration limit to negative infinity, and making the replacement $\tilde{\rho}_0(t')\to \tilde{\rho}_0(t)$. These steps are justified by the assumption that $\gamma_j \gg g,\Gamma$, so that the integrand decays rapidly to zero compared to the time scale over which  $\PP\tilde{\rho}_0(t)$ changes appreciably. 
	
Equation~\eqref{RedFieldEquation} is then simplified by expanding out the commutators, tracing over the engine qubits, and then transforming back to the Schr\"odinger picture. The resulting master equation decouples the evolutions of the populations and coherences when $\rho_w(t)$ is expressed in the eigenbasis of $B_w$. Since by assumption there is no initial coherence (see Eq.~(22) ), we quote only the result for the populations
	\begin{align}
		\label{rhoPpopulationME}
		\dt{\rho_w} = & \;p_\downarrow \DD[B_w]\rho_w + p_\uparrow \DD[B_w^\dag] \rho_w \notag \\ & -\frac{\Gamma}{2} \left ( \ket{d-1}_w\bra{d-1}\rho_w + \rho_w \ket{d-1}_w\bra{d-1} \right ).
	\end{align}
	Introducing the probability vector ${{\bf q}}$ with elements ${{\bf q}}_n(t) =  \Tr[\rho_w(t) \ket{n}_w\bra{n}]$, we have $\dd {{\bf q}} /\dd t = A {{\bf q}}$, with
	\begin{equation}
	A = \begin{pmatrix}
	-p_\uparrow 	& p_\downarrow 				& 				& 			& 			& & &\\
	p_\uparrow 	& -(p_\uparrow+p_\downarrow) 	& 		&			&			& & &\\
	&					&					& \ddots	&		&  & &\\
	& 					& 				&				&  &-(p_\uparrow + p_\downarrow) & p_\downarrow \\
	& 	&&&& p_\uparrow & -(p_\downarrow + \Gamma) 
	\end{pmatrix}.
	\end{equation}
This is equivalent to equation~\eqref{eqmotionstoch} for the probabilities ${\bf q}_n(t) = q(n,t)$, but with an additional term proportional to $\Gamma$ describing spontaneous decay from the upper level. The forward and backward rates are Laplace-transformed correlation functions of the engine qubits,
	\begin{align}
		\label{backwardProbDef}
		p_\downarrow & = 2 g^2 \int_0^\infty\dd t\; \e^{\ii E_w t} \left \langle \sigma_h(t) \sigma_h^\dag(0) \sigma_c^\dag(t) \sigma_c(0)\right \rangle, \\
		\label{forwardProbDef}
		p_\uparrow & = 2 g^2 \int_0^\infty\dd t\; \e^{-\ii E_w t} \left \langle \sigma^\dag_h(t) \sigma_h(0) \sigma_c(t) \sigma_c^\dag(0)\right \rangle ,
	\end{align}
	where the angle brackets denote an average with respect to $\rho_h\otimes \rho_c$, while the operator time dependence is given by $\sigma_{h,c}(t) = \e^{\LL_0^\dag t} \sigma_{h,c}$, where $\LL_0^\dag$ is the adjoint Liouvillian defined by $\Tr [Q \LL_0 (P) ] = \Tr[ \LL_0^\dag(Q) P]$ for arbitrary operators $P$ and $Q$. Explicitly, we have $\sigma_j(t) = \exp(-\ii E_jt - \gamma_j \mathcal{Z}_j t/2 ) \sigma_j$ for $j=h,c$, implying that
	\begin{align}
		\label{backwardProbExplicit}
		p_\downarrow & = \frac{4 g^2 \e^{-\beta_c E_c}}{\mathcal{Z}_h \mathcal{Z}_c(\gamma_h\mathcal{Z}_h + \gamma_c \mathcal{Z}_c)}, \\
		\label{forwardProbExplicit}
		p_\uparrow & = \frac{4 g^2 \e^{-\beta_h E_h}}{\mathcal{Z}_h \mathcal{Z}_c(\gamma_h\mathcal{Z}_h + \gamma_c \mathcal{Z}_c)}, 
	\end{align}
	from which one readily verifies that $p_\uparrow/p_\downarrow = \e^{-(\beta_h E_h - \beta_c E_c)} = \e^{-\beta_v E_w}$. Self-consistency of the Born-Markov approximation requires that $p_\downarrow,p_\uparrow \ll \gamma_j$.
	
\section{Model-independent limits on thermally run clocks}\label{appg}
	
We have argued that the accuracy of the autonomous clock is constrained by the amount of heat that the clock dissipates as it provides ticks. This was obtained by relating the upwards bias of the ladder's evolution to the ratio of populations of the virtual qubit, which for the two-qubit engine, is found to satisfy~\cite{virtual}
	\begin{equation}\label{biasheat}
	\beta_w E_w = \beta_h E_h - \beta_c E_c.
	\end{equation}
	Multiplying by the spectral width of the ladder, $d-1$, and since ($E_c = E_h + E_w$), we find that
	\begin{equation}
	\beta_w E_\gamma = \beta_h \left( Q_c + E_{\gamma} \right) - \beta_c Q_c.
	\end{equation}
This expression may be intuitively understood as follows. Every time the thermal machine prepares the virtual qubit in the appropriate state that is ready to exchange $E_w$ with the external system, it must also absorb $E_h$ from the hot reservoir, and dissipate $E_c$ to the cold reservoir.
	
This is in fact true not only for the two-qubit engine, but has been shown to be the case for a large class of autonomous quantum thermal machines~\cite{Silva16} (in the weak-coupling regime). That is, while the bias can be tweaked by changing the machine design from the two-qubit engine to more complex constructions, it is always constrained to obey equation~\eqref{biasheat}. Therefore, the trade-off between accuracy and power for the autonomous clocks is a general feature not limited to the model in this paper.
	
On the other hand, it would be interesting to investigate clocks that deviate from weak coupling, as they may be able to outperform stochastic models via the build-up of coherence. Even in the simplest case of the two-qubit engine, there is some build-up of coherence in the subspace of the interaction between engine and ladder, that is maintained as the ladder moves upward. In Ref.~\cite{virtual}, this is observed to prevent the ladder's energy distribution from spreading as much as would be expected from a simply stochastic model, which in turn would lead to a higher accuracy. Clocks that are even more coherent (while not necessarily autonomous) have been observed~\cite{ralph} to spread much less than thermal clocks. The possibility of achieving more accurate clocks via the use of stronger couplings and coherence is thus an important direction for future work.	


\begin{thebibliography}{99}
\bibitem{footrank} For an operational definition of accuracy that is independent of background time, see \cite{rank}.
\bibitem{footnote1} Note that this analysis is based only on the statics of the model, and is thus independent of the details of the thermalisation model for describing the coupling between the qubits and the baths.
\end{thebibliography}
\end{document}